\documentclass[prd,aps,showpacs,superscriptaddress,nofootinbib,notitlepage,floatfix]{revtex4-1}

\usepackage{amsmath,latexsym}
\usepackage{graphics}

\usepackage{epsfig,graphicx,amsfonts}

\RequirePackage{mathrsfs}

\def\eq#1{{Eq.~(\ref{#1})}}
\newcommand{\Le}{\left(}
\newcommand{\Ra}{\right)}

\newcommand{\beq}{\begin{equation}}
\newcommand{\eeq}{\end{equation}}
\newcommand{\beqar}{\begin{eqnarray}}
\newcommand{\eeqar}{\end{eqnarray}}
\newcommand{\D}{\partial}

\newcommand{\om}{\omega}

\usepackage{color}
\newcommand{\revisionZ}[1]{{#1}}
\newcommand{\revision}[1]{{#1}}
\newcommand{\revisionF}[1]{{#1}}

\newcommand{\revisionU}[1]{{#1}}

\begin{document}
\title{ High energy scattering in Einstein-Cartan gravity}

\author{S.~Bondarenko}
\affiliation{Physics Department, Ariel University, Ariel 40700, Israel}

\author{S. Pozdnyakov}
\affiliation{Saint Petersburg State University, Saint Petersburg, Russia}

\author{M.A.Zubkov 
\thanks{on leave of absence from Institute for Theoretical and Experimental Physics, B. Cheremushkinskaya 25, Moscow, 117259, Russia}%
}                     
\affiliation{Physics Department, Ariel University, Ariel 40700, Israel}

%
%
%
%
\begin{abstract}
We consider Riemann-Cartan gravity with minimal Palatini action, which is classically equivalent to Einstein gravity. Following the ideas of L.Lipatov \cite{LipGrav} we propose the effective action for this theory aimed at the description of the high-energy scattering of gravitating particles in the  multi - Regge kinematics. For that purpose we add to the Palatini action  the new terms responsible for the interaction of gravitational quanta with certain collective excitations that replace exchange by multiple gravitational excitations. We propose the heuristic explanation of its particular form based on an analogy to the reggeon field theory of QCD. We argue that Regge kinematics assumes the appearance of an effective two - dimensional model describing the high - energy scattering  similar to that of QCD. Such a model  may be formulated in a way leading to our final effective theory, which contains interaction between the ordinary quanta of spin connection and vielbein with their effective counterparts that pretend to the role of the gravitational reggeons.
\end{abstract}
\maketitle

\section{Introduction}

 In Regge theory scattering amplitude at large energies $\sqrt{s}$ and fixed momentum transfer $\sqrt{-t}$ has the form \cite{Grib}:
\beq\label{Int1}
A_{Regge}(s,t)\,\propto\,s^{1\,+\,\om_{p}(t)}\,,
\eeq
where $p\,=\,\pm\,1$ is the signature of Reggeon with trajectory $\om_{p}(t)$.  In quantum field theory particles may acquire properties of Reggeons when radiative corrections are taken into account
\cite{PReg,BFKL}. In QCD the effective Reggeon field theory
\cite{LipatovEff,LipatovEff1,Our1} describes high energy scattering with multi  - Regge kinematics. In this case we operate with an effective action local in rapidity space, which describes
interactions between physical gluons inside rapidity clusters $(y\,-\,\eta/2,y\,+\,\eta/2)$ with $\eta\,\ll\, \ln(s)$.
The whole rapidity interval is covered by these clusters. Interaction between physical gluons by an exchange by multiple virtual gluons may be described effectively by an exchange by single reggeons.  In order to describe these effective interactions the Lipatov's induced effective action may be used. The resulting theory contains the action of the form
\beq\label{Int2}
S_{eff}\,=\,\int\,d^{4} x\,\Le L_{0}\,+\,L_{ind}\,\Ra\,,
\eeq
Here $L_0$ is bare action while $L_{ind}$ is an extra term. This action reproduces \eq{Int1} for the scattering amplitude and correctly describes production of particles in direct channels in quasi-multi-Regge kinematics,
it is highly non-linear and gauge invariant, see \cite{EffAct,Fadin,Our2}.

 So far, this effective action formalism was applied to the calculations of various QCD processes, see \cite{EffAct,Fadin,Our2,Nefedov}. An alternative approachs to high energy scattering are based on the concept of Color Glass Condensate and on the Wilson line formalism of I.Balitsky \cite{CGC,Balit,BK,Hetch,CGC1}. An attempt to apply  similar approach to the high energy gravitational scattering \cite{LipGrav} encountered certain problems.
The reggeization of gravition, similar to the gluon's reggeization, has been discussed a while ago \cite{GravReg}, see also the recent papers \cite{GravReg1}. Certain effective gravitational actions were applied to the description of  scattering \cite{AmAct}. Nevertheless, to the best of our knowledge, Regge effective theory for the high energy scattering in quantum gravity has not been completed (as well as the quantum gravity itself, though).

 Construction of \cite{LipGrav} operates with the conventional general relativity, and the heuristic approach to the reggeization has been proposed. The necessary terms of effective action have been considered in \cite{LipGrav}, and are obtained up to the second order in the weak field approximation. Technically this approach experiences difficulties related to the lack of the machinery similar to that of the Wilson lines in  QCD.  In the present paper we suggest a way to resolve this difficulty based on the consideration of Einstein gravity in first order formalism with independent vielbein and spin connection.

We begin from the Riemann-Cartan-Einstein (RCE) gravity with minimal Palatini action, which is classically equivalent to Einstein gravity. This theory, however, is initially formulated as a gauge field theory with $so(3,1)$ gauge group. Therefore, the formalism of \cite{LipatovEff,LipatovEff1,Our1} can be directly transferred to the framework of RCE gravity.  We formulate the effective Regge theory for this model. Wilson lines of \cite{LipatovEff,LipatovEff1,Our1} have their  counterparts in the RCE gravity. The constructed effective action is to be invariant with respect to the corresponding gauge transformations. The price for the methodological simplification is the complication of calculations. In the RCE gravity there are two independent fields: spin connection $\om_{\mu}$ and vielbein $e_{\mu}^{a}$. An attempt to reduce the theory to the one expressed through the field of metric assumes (in a certain approximation) that the equations of motion with respect to connection are to be solved. Then connection is expressed through vielbein. The vielbein, in turn, is related to the field of metric. The final form of effective action contains metric field only.
After this reduction it is really hard to recognize in the final expressions the initial gauge invariant ones. This is the reason for the complicated form of expressions proposed in \cite{LipGrav}.

The paper is organized as follows. First of all, we repeat the dimensional reduction procedure similar to the one of
\cite{Our3}.  This is done in Section 2. Due to kinematics of high energy scattering the effective theory in the center of inertia reference frame has to be two - dimensional.  The effective theory operates with the gauge invariant quantities - the Wilson lines of $SO(3,1)$ group (taken along the light cone), and the invariant translation along the light cone given by vielbein. The coefficients in effective action may be fixed comparing its form with that of bare action in the ultrahigh energy limit, when gauge field may be thought of as constant during the time of collision. In Section 3 we use Hubbard - Stratonovich transformation in order to bring the effective two - dimensional theory to the four - dimensional form with the auxilary fields. These auxiliary fields interacts with ordinary vierbein and spin connection. Exchange by one quantum of such fields replaces the exchange by multiple gravitational excitations.  This procedure, however, remains ambiguous. It admits an addition of extra terms that disappear in ultra - relativistic regime. This way in Section 4 we modify the obtained effective action in order to make it symmetric under the time reversal. We also analyse briefly the classical equations of motion for the gravitational excitations in the presence of the mentioned effective excitations.
In Section 5 we end with the conclusions.

\section{The effective $2D$ model for the high energy scattering}

\label{2D}

Here we present the dimensional reduction procedure similar to that of
\cite{Our3}. The main idea is that due to kinematics of high energy scattering the effective theory in the center of inertia reference frame of the two colliding gravitating particles has to be two - dimensional (coordinates of time and direction of collision are compactified).  The effective theory operates with the gauge invariant quantities - the Wilson lines of $SO(3,1)$ group (taken along the light cone), and the invariant translation along the light cone given by vielbein. The reason for this is that the high energy particle moving close to the light cone interacts with the gravitational field via these two quantities.

We choose the following dynamical variables.
The $2D$ components of the gauge field $\omega_i(x^+,x^-,\vec{x}_\bot)$ $\in$ $so(3,1)$ ($i=2,3$) and the extra $so(3,1)$ fields $\omega_+(x^+,x^-,\vec{x}_\bot)$, $\omega_-(x^+,x^-,\vec{x}_\bot)$ that originate from the light cone components of the original $4D$ spin connection. It is assumed that metric is close to the flat one, and by the light cone components we understand the light cone components in the unperturbed metric of Minkowski space. This means, in particular, that the world trajectories of light only approximately are along this "light cone".
We start from the Palatini action
\begin{eqnarray}\label{Sec1}
S&=&-{m_P^2}{\rm Tr}\,\int d^4 x \, |e| e^{\mu}_a e^{\nu}_b G^{ab}_{\mu\nu}
\end{eqnarray}
where $G_{\mu \nu}$ is the field strength, $m_P$ is the Plank mass. In order to write down the classical equations of motion that are the part of perturbative expansion (see below) we prefer to use the representation  written in terms of the vielbein $e^a_\mu$ without using its inverse $e^\mu_a$. The corresponding form of Palatini action is:
\beq\label{AddF1}
S^{0}[\omega]\,=\,\frac{m_{P}^{2}}{2}\,\int\,d^4 x\,\varepsilon^{\mu \nu \rho \sigma}\,\varepsilon_{a b c d}\,e_{\rho}^{c}\,e_{\sigma}^{d}\,\Le D_{\mu} \om_{\nu}^{a b} \Ra\,
\eeq

We chose the $4$ - vector  $n^\mu = (1,1,0,0)$ and denote $\vec{n} = (1,0,0)$. The light - cone notations are:
\beq\label{Sec2}
x^- = x_+ = \frac{x^0 - x^1}{\sqrt{2}},\,\, \quad x^+ = x_- = \frac{x^0 + x^1}{\sqrt{2}},\,\, \vec{x}_\bot = (0,0,x^2,x^3)
\eeq
As it was mentioned above, it is assumed that metric field $g_{\mu\nu}$ is close to its Minkowski form ${\rm diag}\,(1,-1,-1,-1)$. Then vielbein $e_a^\mu$ may be chosen close to ${\rm diag}\,(1,1,1,1)$ by an appropriate $so(3,1)$ gauge transformation. Then nonzero light - cone components of unperturbed $g$ are $g^{+-}=g^{-+}=-g^{22}=-g^{33}=1$, while nonzero light - cone components of unperturbed vielbein $e_a^\mu$ are $e_0^+=e_1^+=1/\sqrt{2}, e_0^-=-e_1^-=1/\sqrt{2}, e^2_2=e^3_3=1$.
We rewrite action in the light - cone coordinates as follows
\begin{eqnarray}\label{Gs}
S&=&-{m_P^2}{}{\rm Tr}\,\int d^4 x \, |e| e^i_a e^j_b G^{ab}_{ij} -{m_P^2}{}{\rm Tr}\,\int d^4 x \, |e| e^\alpha_a e^\beta_b G^{ab}_{\alpha\beta}-2{m_P^2}{}\,\int d^4 x \, |e| e^\alpha_a e_b^i G_{\alpha i}\nonumber\\&&  i,j,k,l = 2,3; \alpha,\beta,\gamma,\delta = \pm
\end{eqnarray}
or, equivalently, as
\begin{eqnarray}\label{Gs}
S&=&\frac{m_P^2}{4}{\rm Tr}\,\int d^4 x \, \,\varepsilon^{ij \rho \sigma}\,\varepsilon_{a b c d}\,e_{\rho}^{c}\,e_{\sigma}^{d}\, G^{ab}_{ij} +\frac{m_P^2}{4}{\rm Tr}\,\int d^4 x \, \,\varepsilon^{\alpha \beta \rho \sigma}\,\varepsilon_{a b c d}\,e_{\rho}^{c}\,e_{\sigma}^{d}\, G^{ab}_{\alpha\beta}+\frac{m_P^2}{2}\,\int d^4 x \, \,\varepsilon^{\alpha i \rho \sigma}\,\varepsilon_{a b c d}\,e_{\rho}^{c}\,e_{\sigma}^{d}\, G_{\alpha i}\nonumber\\&&  i,j,k,l = 2,3; \alpha,\beta,\gamma,\delta = \pm
\end{eqnarray}

The direct dimensional reduction $3+1$D $\to 2$D works as follows. We should consider the $3+1$ D space - time, in which the coordinates $x^\pm$ belong to the interval $0<x^\pm<\Delta$, where $\Delta = \frac{1}{\Lambda}$ and $\Lambda$ is the parameter of the dimension of mass, which is supposed to be much larger than the typical energies of the virtual gravitational excitations radiated and absorbed during the scattering process.

In the effective $2D$ model instead of the algebra elements $\omega_\pm$ we operate with the group elements of the original $4D$ theory. Wilson lines taken in vector representation are:
\revisionF{\begin{eqnarray}
{\cal T}_+(x_\bot) &=&P\,  {\rm exp}\,\Big( \int_{0}^{\Delta} \omega^-(\vec{x}_\bot)d \bar{x}^+ \Big) \nonumber\\
{\cal T}_-(x_\bot) &=& P\, {\rm exp}\,\Big(  \int_{0}^{\Delta} \omega^+(\vec{x}_\bot)d\bar{x}^- \Big)\label{E8}
\end{eqnarray}}
 Eq. (\ref{E8}) represents elements of Lorentz group while the following expressions represent elements of the group of translations:
\begin{eqnarray}
{\cal F}^b_+(x_\bot) &=&   \int_{0}^{\Delta} e^b_-(\vec{x}_\bot)d \bar{x}^+ \nonumber\\
{\cal F}^b_-(x_\bot) &=&   \int_{0}^{\Delta} e^b_+(\vec{x}_\bot)d\bar{x}^- \label{F8}
\end{eqnarray}
$\cal F$ and $\cal T$ together form the elements of Poincare group.
  We denote
$$
q^- = q_+ = \frac{q_0 + q_1}{\sqrt{2}}= \frac{q^0 - q^1}{\sqrt{2}}$$ $$ q^+ = q_- = \frac{q_0 - q_1}{\sqrt{2}}=\frac{q^0 + q^1}{\sqrt{2}}$$
$$ \vec{q}_\bot = (0,0,q_3,q_4)
$$
With these notations
$$
q_\mu x^\mu = q^0x^0 - \vec{q}\vec{x} = q_+ x^+ + q_- x^- - (\vec{q}_\bot \vec{x}_\bot)
$$
For the dimensional reasons the possible effective action of the mentioned $2D$ theory may be written as
\begin{eqnarray}
S_{+-} &=&{m_P^2}\int d^2 \vec{x}_\bot \, \Big(\varepsilon^{+- i \rho }\,\varepsilon_{a b c d}\,{e}_{\rho}^{d}\,{\cal F}_{-}^{c}\,D_i {\cal T}^{ab}_+  + \varepsilon^{-+ i \rho }\,\varepsilon_{a b c d}\,{e}_{\rho}^{d}\,{\cal F}_{+}^{c}\, D_i {\cal T}^{ab}_-  \Big)\label{S+-}
\end{eqnarray}
Here $a,b,c,d,\rho, \sigma = 0,1,2,3$ while $i = 1,2$. The coefficient in front of this action is to be fixed comparing it to the Palatini action in the ultra - high energy regime, when $\omega$ does not depend on $x^\pm$. The dimensional reduction also assumes smallness of the product $\omega_\pm \Delta$, therefore Eq. (\ref{Gs}) may be rewritten with the help of  ${\cal T}_\pm$.
 \revisionF{The integration over $x^\pm$ is effectively reduced to the integration over the intervals of lengths $1/\Lambda$ because we describe the high energy scattering in the center of inertia reference frame, and the scattered particles have such a large energy that during the collision the fields $\omega^\pm,\omega_a$ have not time to change. $1/\Lambda$ is of the order of the collision time. The uncertainty relation assumes $\Lambda \sim \sqrt{s}$. The discussion of the  different approaches to the description of this dimensional reduction may be found in \cite{Verlinde}.}

Because of the large value of $\Lambda$ we neglect the dependence of $\omega$ on $x^\pm$.
The value of $\Lambda$ is of the order of magnitude of $\sqrt{s}$ while the typical momentum transfer is $\sqrt{-t} \ll \sqrt{s}$. Therefore, we neglect in Eq. (\ref{Gs}) the terms containing $\Lambda^2$ in the denominator. As a result we are left with
\beq\label{Gs3}
S\,=\,m_{P}^{2}\,\int\,d^4 x\,\varepsilon^{i - + \sigma}\,\varepsilon_{a b c d}\,e_{+}^{c}\,e_{\sigma}^{d}\,\Le D_{i} \om_{-}^{a b} \Ra\,+\,
\,m_{P}^{2}\,\int\,d^4 x\,\varepsilon^{i + - \sigma}\,\varepsilon_{a b c d}\,e_{-}^{c}\,e_{\sigma}^{d}\,\Le D_{i} \om_{+}^{a b} \Ra\,.
\eeq
where $D_i$ is the covariant derivative with respect to the $2D$ gauge field.

\revisionU{The equivalence between \eq{S+-} and \eq{Gs3}, when $\omega, e$ do not depend on $x^\pm$ while $\omega_\pm \Delta \ll 1$, can be demonstrated by the following calculation. We have
for the first term in \eq{S+-} to the leading order precision:
\begin{eqnarray}\label{Sec4}
&&  m_P^2 \int d^2 \vec{x}_\bot \varepsilon^{+- i \rho }\,\varepsilon_{a b c d}\,{e}_{\rho}^{d}\, D_i {\cal T}^{ab}_+ {\cal F}^c_{-}  = m_P^2 \int d^2 \vec{x}_\bot  \varepsilon^{+- i \rho }\,\varepsilon_{a b c d}\,{e}_{\rho}^{d}\, D_i e^{  \int_{0}^{\Delta} \omega^-(\vec{x}_\bot)d {x}^+ }  \int_{0}^{\Delta} e^c_-(\vec{x}_\bot) d {x}^- =\nonumber\\
&&=m_P^2\int d^2 \vec{x}_\bot  \varepsilon^{+- i \rho }\,\varepsilon_{a b c d}\,\int_{0}^{\Delta} d {x}^+  \int_{0}^{\Delta} d {x}^- \,\,{e}_{\rho}^{d} D_i   \omega^{-,ab}(\vec{x}_\bot)  e^c_-(\vec{x}_\bot)
\nonumber\\&&=m_P^2\int d^2 \vec{x}_\bot  d {x}^+  d {x}^-   \varepsilon^{+- i \rho }\,\varepsilon_{a b c d}\,{e}_{\rho}^{d} D_i    \omega^{-,ab}  e^c_-
\end{eqnarray}
Calculating similarly the other term of \eq{S+-} and summing them together we come finally to expression of Eq. (\ref{Gs3}).}

This form of the action may be used in the framework of the $3+1$ D theory for the bare description of the high energy scattering at $s/(-t) \gg 1$, and by $\cal T$ we should understand
\begin{eqnarray}
{\cal T}_+(x_\bot) &=& {\cal T}_+(x^-,x_\bot)\Big|_{x^-=0}, \quad  {\cal T}_+(x^-,x_\bot) =  P\, {\rm exp}\,\Big(  \int_{-\infty}^{\infty} \omega^-(\bar{x}^+,x^-,\vec{x}_\bot)d \bar{x}^+ \Big)\nonumber\\
{\cal T}_-(x_\bot) &=& {\cal T}_-(x^+,x_\bot)\Big|_{x^+=0}, \quad  {\cal T}_-(x^+,x_\bot) =  P\, {\rm exp}\,\Big(  \int_{-\infty}^{\infty} \omega^+(x^+,\bar{x}^-,\vec{x}_\bot)d\bar{x}^- \Big)\label{E13}
\end{eqnarray}
Parallel transporters in vector representation are denoted by
\begin{eqnarray}
\Big[{\Omega}_+(x^+,x^-,x_\bot)\Big]^a_{.b} &=& \Big[P\, {\rm exp}\,\Big(  \int_{-\infty}^{x^+} \omega^-(\bar{x}^+,x^-,\vec{x}_\bot)d \bar{x}^+ \Big)\Big]^a_{.b} \nonumber\\
\Big[{\Omega}_-(x^+,x^-,x_\bot)\Big]^a_{.b} &=&\Big[P\, {\rm exp}\,\Big(  \int_{-\infty}^{x^-} \omega^+(x^+,\bar{x}^-,\vec{x}_\bot)d\bar{x}^- \Big)\Big]^a_{.b} \label{E14}
\end{eqnarray}
$SO(3,1)$ indices are lowered and lifted using metric tensor $\eta_{ab}$ of Minkowski space: $\Big[{\Omega}_\pm\Big]^a_{.b} = \Omega_\pm^{a\bar{b}}\eta_{\bar{b} b}$. In addition to the vierbein $e_\mu^a(x)$ that is transformed via the $SO(3,1)$ gauge transformation (it acts on index $a$) localized at $x$ we define quantity transformed via gauge transformation at (minus) infinity:
\begin{eqnarray}
{ E}_+^{c}(x^+,x^-,x_\bot) &=&  e^a_+({x}^+,x^-,\vec{x}_\bot)  \Omega^{bc}_+({x}^+,x^-,x_\bot) \eta_{ab} \nonumber\\
{ E}_-^{c}(x^+,x^-,x_\bot) &=&    e^a_-(x^+,{x}^-,\vec{x}_\bot)\Omega^{bc}_-(x^+,{x}^-,x_\bot) \eta_{ab} \label{E15_}
\end{eqnarray}
Besides, we define the translational parallel transporters along the  "light cone":
 \begin{eqnarray}
 {\cal F}_{+}^c(x_\bot) &=& {\cal F}_{+}^c(x^-,x_\bot) \Big|_{x^-=0}, \quad {\cal F}_{+}^c(x^-,x_\bot) = \int_{-\infty}^{\infty} { E}^c_+(\bar{x}^+,x^-,\vec{x}_\bot)d \bar{x}^+ \nonumber\\
{\cal F}_{-}^c(x_\bot) &=&  {\cal F}_{-}^c(x^+,x_\bot) \Big|_{x^+=0}, \quad {\cal F}_{-}^c(x^+,x_\bot) =  \int_{-\infty}^{\infty} {E}_-^c(x^+,\bar{x}^-,\vec{x}_\bot)d\bar{x}^- \label{E15}
\end{eqnarray}

\revisionF{This is the basic idea of the present paper: we consider the $3+1$ D theory, and take as a first approximation to the effective action the  action of the reduced $2D$ theory, where instead of the group elements of Eq. (\ref{E8}) the Wilson lines of Eq. (\ref{E13}) are substituted. Next, we take into account an extra exchange by the virtual soft gravitons and thus come to the effective action a la Lipatov.}

Let us recall that the considered theory is gauge theory of $SO(3,1)$ group and also the gauge theory of the group of translations.
Basing on an analogy to QCD we then suppose that
in the center of mass reference frame the scattering amplitude of two gravitating particles (for the case when $s \gg |t|$,  in eikonal approximation) contains the leading factor
\begin{equation}
D(q_\bot) \sim \langle \int d^2 x_\bot e^{iq_\bot x_\bot} \Big[ {\cal F}_{+}(x_\bot)\oplus  ({\cal T}_+(x_\bot)-1)\Big]\otimes \Big[ {\cal F}_{-}(0) \oplus ( {\cal T}_-(0)-1) \Big] {\cal R}(x_\bot) \rangle_{e,\omega}\label{Scattering}
\end{equation}
Here $q_\bot$ is the momentum transfer orthogonal to the axis of collision while averaging is over the gravitational fields $e$ and $\omega$. This expression contains the parallel transporters (both in the group of translations and in the SO(3,1) group) along the light cone. These parallel transporters should appear in Schwinger representation of the theory as the sum over trajectories of the colliding particles. In Eikonal approximation the sum over trajectories is reduced to the trajectories along the light cone. The appearance of the parallel transporters follows from gauge invariance. It is supposed that like in the case of QCD they give dominant contribution compared to the form - factor ${\cal R}$ that accounts for the other contributions. In Eq. (\ref{Scattering}) we use symbols $\otimes$ and $\oplus$ to represent symbolically that various combinations like ${\cal T} {\cal T}$, ${\cal T} {\cal F}$, ${\cal F}{\cal F}$ are present. The corresponding indices are contracted with the indices of ${\cal R}$ and with the indices corresponding to the states of the colliding particles.


\section{Candidates to the role of gravitational reggeons}
\label{Lipatov}

Below we will show, that the effective action similar to the one proposed by Lipatov appears, when we consider the ordinary Palatini action $S^{0}[\omega]$ supplemented by the $2D$ model introduced above. We justify this procedure as follows. Bare effective $2D$ action for the high energy scattering is the action of the deduced above model. We consider the fields ${\cal T}_{\pm}, {\cal F}_\pm$ as candidates to the role of bare gravitational reggeons. Exchange by these fields between the gravitating particles during the high energy scattering substitutes the exchange by multiple ordinary gravitational excitations. This occurs because the scattering amplitude of two particles in high energy scattering is proportional to the correlators of the parallel transporters along the light cone (recall that ${\cal T}_{\pm}$ is the parallel transporter of $SO(3,1)$ group while  $ {\cal F}_\pm$ is the translational parallel transporter).

Interaction of quantities ${\cal T}_{\pm}, {\cal F}_\pm$ with the ordinary gravitational excitations (with Palatini action) results in modification of the bare propagators. We propose the hypothesis that the corresponding dressed propagators are actually the propagators of proper gravitational reggeons. The direct check of this hypothesis requires the investigation  of the asymptotic form of the corresponding propagators as well as the other conditions of reggeization. This remains out of the scope of the present paper.

Here we present the construction similar to that of QCD that has led to the Lipatov effective action for the reggeons.
We obtain corrections to the above mentioned effective two dimensional action due to interactions with the ordinary virtual excitations of vielbein and spin connection. In order to take into account these corrections we add the action of pure gravity $S^{(0)}[\omega]$ to the action of the $2D$  model of Eq. (\ref{S+-}). At the same time in the latter we substitute the covariant derivative $D_i$ by the ordinary one, which means that we fix the boundary conditions with the vanishing values of  $\omega_i(\pm \infty, x^-, \vec{x}_\bot)$ and $\omega_i(x^+,\pm \infty, \vec{x}_\bot)$ for $i = 2,3$.

  Since the effective two - dimensional model appears in the dimensional reduction of gravity, it is already included into the action $S^{(0)}$.
Therefore, we cannot simply add one to another. We should add an additional prescription, which allows to avoid overcounting in the further perturbative calculations. The prescription for this avoiding is actually very simple. Its essence is the understanding that our candidates to the role of the reggeons appear as the classical solutions of the equations of motion. Namely, we will see that the theory with the action that consists of Eq. (\ref{S+-}) and $S^{(0)}$ may be rewritten in terms of the auxiliary fields $\cal A$ and $\cal E$ that are in certain sense dual to spin connection and vierbein respectively. In the presence of nonzero $\cal A$ and $\cal E$ the classical solution for the ordinary spin connection $\omega$ and $e$ is highly nontrivial, it depends in a complicated way on $\cal A$ and $\cal E$. The \revisionZ{most natural way to construct the perturbation theory assumes that the perturbations around this classical solution are considered. However, this procedure would lead to the double counting the degrees of freedom. Therefore, in order to avoid the overcounting we should build the perturbation theory around $\omega=0$ and flat vierbein $e$ corresponding to Minkowski space.}

We consider the two sets of variables: the non - local $2D$ fields ${\cal T}_\pm (x_\bot)$, ${\cal F}_\pm (x_\bot)$ and the local fields $\omega(x^+,x^-,x_\bot)$, $e(x^+,x^-,x_\bot)$. We suppose, that the local fields tend to the trivial values at infinity. This means, in particular, that $e_i^a $ entering Eq. (\ref{S+-}) is trivial being defined at $(x_+,x_-) = (- \infty,0)$ and $(x_+,x_-) = (0,- \infty)$. Correspondingly, the transverse covariant derivative becomes the usual derivative.
Thus we consider gravity with the following modified action
\begin{eqnarray}\label{Ch1}
S[\omega]  &=& S^{0}[\omega]  \revisionU{+} m_P^2  \int d^2 x_\bot \, \varepsilon^{i - + \sigma}\,\varepsilon_{a b c d}\, \delta_\sigma^d\, \Big( \partial_i {\cal T}^{ab}_-(x_\bot)
 {\cal F}^c_{+} (x_\bot)+ \partial_i {\cal T}^{ab}_+ (x_\bot){\cal F}^c_{-}(x_\bot)\Big)
\end{eqnarray}
In the following we will use matrices $\lambda^a$ that are the generators of the $SO(3,1)$ group in adjoint representation. (The matrix proportional to unity is added to the set of the generators.) We normalize those matrices of the generators in adjoint representation in such a way that ${\rm Tr}\,\lambda^a\,\lambda^b = \delta^{ab}$. Next, we introduce the new variable ${\cal A}^\pm={\cal A}_\mp = {\cal A}_\mp^a \lambda^a \in so(3,1)\oplus so(3,1)$. Here the components ${\cal A}_\mp^a$ are the complex numbers.

Using expressions of Eqs. (\ref{E15_}), (\ref{E15}) in the following we omit the $SO(3,1)$ indices for simplicity.
We assume that at $x_- = \pm \infty$ and at $x_+ = \pm \infty$ the vielbein is trivial $e^a_\mu = \delta^a_\mu$. Considering $SO(3,1)$ gauge transformations $h$ we assume, that $h \to 1$ at $x^\pm \to \pm \infty$. The original theory is invariant under reparametrizations $x \to \tilde{x}$ such that $\tilde{x} = x $ at $x^\pm \to \pm \infty$. However, in the theory with action of Eq. (\ref{Ch1}) invariance under reparametrizations is broken by the choice of the straight lines going along the axis $x^+$ and $x^-$. To demonstrate this it is enough to consider modification under the reparametrizations of ${\cal T}^{ab}_\pm$ and ${\cal F}_{\pm,b}$. To do this we write
\begin{eqnarray}\label{Ch21}
{\cal T}_+(x^-,x_\bot) &=&P\, {\rm exp}\,\Big(  \int_{C_+} \omega_\mu(x)d x^\mu \Big)\nonumber\\
{\cal T}_-(x^+,x_\bot) &=& P\, {\rm exp}\,\Big(  \int_{C_-} \omega_\mu(x)dx^\mu \Big)
\end{eqnarray}
Here contours $C_\pm$ go along the line of $x_\pm$ from minus infinity to plus infinity. Reparametrization results in $x\to \tilde{x}(x)$,  $dx^\mu = \frac{\partial x^\mu}{\partial \tilde{x}^\nu} d\tilde{x}^\nu$, and $\omega_\mu(x) = \frac{\partial \tilde{x}^\nu}{\partial {x}^\mu} \tilde{\omega}_\nu(\tilde{x})$:
\begin{eqnarray}\label{Ch22}
{\cal T}_+(x^-,x_\bot) &=&P\, {\rm exp}\,\Big(  \int_{C_+} \frac{\partial \tilde{x}^\nu}{\partial {x}^\mu} \tilde{\omega}_\nu(\tilde{x}(x))\frac{\partial x^\mu}{\partial \tilde{x}^\rho} d\tilde{x}^\rho(x) \Big) = P\, {\rm exp}\,\Big(  \int_{\tilde{x}(C_+)}  \tilde{\omega}_\nu(\tilde{x}) d\tilde{x}^\nu \Big) \nonumber\\
{\cal T}_-(x^+,x_\bot) &=& P\, {\rm exp}\,\Big(  \int_{\tilde{x}(C_-)} \tilde{\omega}_\mu(\tilde{x})d\tilde{x}^\mu \Big)
\end{eqnarray}
We see, that reparametrizations result in modification of the form of $C\pm$: instead of straight lines we have certain curves given by functions $\tilde{x}(x)$. The similar modification appears in expressions for ${\cal F}_\pm$.
\revisionU{We also introduce the following notations for some operators of interest:
\begin{eqnarray}\label{Ch4}
{\Phi}_+(x^-,x_\bot) &=&P\, {\rm exp}\,\Big(  \int_{-\infty}^\infty \omega^-(x^+,x^-,\vec{x}_\bot)d x^+ \Big) -1  \nonumber\\
{\Phi}_-(x^+,x_\bot) &=& P\, {\rm exp}\,\Big(  \int_{-\infty}^\infty \omega^+(x^+,x^-,\vec{x}_\bot)dx^- \Big)-1\nonumber
\end{eqnarray}}

The partition function may be represented as follows
\begin{eqnarray}\label{Ch6_}
Z  &=&\, \int D\omega \, {\rm exp}\,\Big(i S^0[\omega] +
\nonumber \\
&\revisionU{+i}&{m_P^2}{} \int d^2 x_\bot  \varepsilon^{i - + \sigma}\,\varepsilon_{a b c d}\, \Big[\Le  \,\int_{-\infty}^{\infty}\,d x^+\,e^c_+(x^+,0,x_\bot)\,\Omega_+(x^+,0,x_\bot)\,\Ra\, \delta^d_\sigma \partial_i
\,\Le P\,e^{ \,\int_{-\infty}^{\infty}\,d x^-\, \,\omega^+(0,x^-,x_\bot)\,}\,-\,1\,\Ra ^{ab}\,+
 \nonumber\\
&&\,+\Le  \,\int_{-\infty}^{\infty}\,d x^-\, e^c_-(0,x^{-},x_\bot)\,\Omega_-(0,x^-,x_\bot)\,\Ra\, \delta^d_\sigma \partial_i
\,\Le P\,e^{ \,\int_{-\infty}^{\infty}\,d x^+\, \,\omega^-(x^+,0,x_\bot)\,}\,-\,1\,\Ra ^{ab}\, \Big] =
\nonumber\\
&=& {\rm const}\, \int D\omega D{\cal A} D{\cal E} \, {\rm exp}\,\Big(i S^0[\omega] - \,
\nonumber\\&&
   \revisionU{-i}m_P^2 \int d^4 x \varepsilon^{i - + \sigma}\,\varepsilon_{a b c d}\,  {\rm Tr}\, \Big[{{\cal E}^c_{+}}{} - {\delta(x^+)}{} {\cal F}^c_{+}(x^-,x_\bot) \, \Big] \delta^d_\sigma \partial_i
 \Big[  {{\cal A}^{ab}_-(x)}{}-  \Le {\cal T}_-(x^+,x_\bot)-\,1\,\Ra^{ab}\,{\delta(x^-)}{} \Big] -
\nonumber\\&&
   \revisionU{-i}m_P^2 \int d^4 x  \varepsilon^{i - + \sigma}\,\varepsilon_{a b c d}\, {\rm Tr}\, \Big[{{\cal E}^c_{-}}{} - {\delta(x^-)}{} {\cal F}^c_{-}(x^+,x_\bot) \, \Big] \delta^d_\sigma \partial_i
\Big[  {{\cal A}^{ab}_+(x)}{}-  \Le {\cal T}_+(x^-,x_\bot)-\,1\,\Ra^{ab}\,{\delta(x^+)}{} \Big]+ \nonumber\\&&
\revisionU{+i}{m_P^2}{} \int d^2 x_\bot  \varepsilon^{i - + \sigma}\,\varepsilon_{a b c d}\, \Big[\Le  \,\int_{-\infty}^{\infty}\,d x^+\, e^c_+(x^+,0,x_\bot)\,\Omega_+(x^+,0,x_\bot)\,\Ra\, \delta^d_\sigma \partial_i
\,\Le P\,e^{ \,\int_{-\infty}^{\infty}\,d x^-\, \,\omega^+(0,x^-,x_\bot)\,}\,-\,1\,\Ra ^{ab}\, -
\nonumber \\
&-&\Le  \,\int_{-\infty}^{\infty}\,d x^+\,e^c_-(x^+,0,x_\bot)\, \Omega_-(0,x^-,x_\bot) \,\Ra\, \delta^d_\sigma \partial_i
\,\Le P\,e^{ \,\int_{-\infty}^{\infty}\,d x^+\, \,\omega^-(x^+,0,x_\bot)\,}\,-\,1\,\Ra ^{ab}\, \Big] \Big)=
 \nonumber\\
&=&{\rm const}\, \int D\omega D{\cal A} D{\cal E}\, {\rm exp}\,\Big(i S_{eff}[\omega,e,{\cal A},{\cal E}]\Big)\label{Z1}
\end{eqnarray}
Here $S_{eff}=S^0[\om] + S_{ind}$ is an effective action for the interaction between auxiliary fields $\cal A$ and $\cal E$ and virtual gravitons:
\begin{eqnarray}\label{Ch7}
&& S_{eff}[\omega,e,{\cal A},{\cal E}]  = S^{0}[\omega,e]
\revisionU{-}{m_P^2} {\rm Tr}\,\int d^4 x \varepsilon^{i - + \sigma}\,\varepsilon_{a b c d}\, \delta_\sigma^d\,  {\cal E}^c_{+}\, \partial_{\bot,i} {\cal A}_-^{ab}\revisionU{-}{m_P^2} {\rm Tr}\,\int d^4 x \varepsilon^{i - + \sigma}\,\varepsilon_{a b c d}\, \delta_\sigma^d\,  {\cal E}^c_{-}\,\partial_{\bot,i} {\cal A}_+^{ab}+
\nonumber\\&&
   \revisionU{+}m_P^2\int d^4 x  \varepsilon^{i - + \sigma}\,\varepsilon_{a b c d}\, \delta_\sigma^d\,  \delta(x^+)  \partial_{\bot,i}\Phi_+^{ab}  {\cal E}^c_{-}(x)
 \revisionU{+}m_P^2\int d^4 x  \varepsilon^{i - + \sigma}\,\varepsilon_{a b c d}\, \delta_\sigma^d\,  \delta(x^-) \, \partial_{\bot,i}\Phi^{ab}_- {\cal E}^c_{+}(x) +\nonumber\\&&
   \revisionU{+}m_P^2\int d^4 x \varepsilon^{i - + \sigma}\,\varepsilon_{a b c d}\, \delta_\sigma^d\,  \delta(x^+) \partial_{\bot,i}{\cal A}_+^{ab}  {\cal F}^c_{-}(x)
 \revisionU{+}m_P^2\int d^4 x  \varepsilon^{i - + \sigma}\,\varepsilon_{a b c d}\, \delta_\sigma^d\,  \delta(x^-) \, \partial_{\bot,i}{\cal A}^{ab}_- {\cal F}^c_{+}(x)
\end{eqnarray}
\revisionU{which can be also represented as follows:}
\begin{eqnarray}\label{Ch8}
&&S_{eff}[A,e,{\cal A},{\cal E}]  = S^{0}[A,e]
\revisionU{-}{m_P^2} {\rm Tr}\,\int d^4 x \varepsilon^{i - + \sigma}\,\varepsilon_{a b c d}\, \delta_\sigma^d\,  {\cal E}^c_{+}(x^+,x^-,x_\bot)\, \partial_{\bot, i} {\cal A}^{ab}_-(x^+,x^-,x_\bot)-
\nonumber\\&&
\revisionU{-}{m_P^2} {\rm Tr}\,\int d^4 x \varepsilon^{i - + \sigma}\,\varepsilon_{a b c d}\, \delta_\sigma^d\,  {\cal E}^c_{-}(x^+,x^-,x_\bot)\,\partial_{\bot, i} {\cal A}^{ab}_+(x^+,x^-,x_\bot)+
\nonumber\\&& \nonumber\\&&
  \revisionU{+}m_P^2 \, \int d^4 x \varepsilon^{i - + \sigma}\,\varepsilon_{a b c d}\, \delta_\sigma^d\,  \partial_{x^+}    \partial_{\bot,i}\Bigl( P\,e^{\imath \,\int_{-\infty}^{x^+}\,d \bar{x}^+\, \,\omega^-(\bar{x}^+,x^-,x_\bot)\,}-P\,e^{\imath \,\int^{\infty}_{x^+}\,d \bar{x}^+\, \,\omega^-(\bar{x}^+,x^-,x_\bot)\,}\,\Bigr)^{ab}\, {\cal E}^c_{-}(0,x^-,\vec{x}_\bot)+
\nonumber\\&&
  \revisionU{+}m_P^2 \, \int d^4 x \varepsilon^{i - + \sigma}\,\varepsilon_{a b c d}\, \delta_\sigma^d\, \partial_{x^-} \partial_{\bot,i}  \,\Bigl( P\,e^{\imath \,\int_{-\infty}^{x^-}\,d \bar{x}^-\, \,\omega^+(x^+,\bar{x}^-, x_\bot)}-P\,e^{\imath \,\int^{\infty}_{x^-}\,d \bar{x}^-\, \,\omega^+(x^+,\bar{x}^-, x_\bot)\,}\,\Bigr)^{ab}\,  {\cal E}^c_{+}(x^+,0,\vec{x}_\bot)+
	\nonumber\\&&
  \revisionU{+}m_P^2 \, \int d^4 x \varepsilon^{i - + \sigma}\,\varepsilon_{a b c d}\, \delta_\sigma^d\,  \partial_{x^+}  \Bigl( \int_{-\infty}^{x^+}\,d \bar{x}^+\, \,E^c_+(\bar{x}^+,x^-,x_\bot)\,-\int^{\infty}_{x^+}\,d \bar{x}^+\, \,E^c_+(\bar{x}^+,x^-,x_\bot)\,\,\Bigr)^{ab}\, \partial_{\bot,i} {\cal A}_{-}^{ab}(0,x^-,\vec{x}_\bot)+\nonumber\\&&
  \revisionU{+}m_P^2 \, \int d^4 x \varepsilon^{i - + \sigma}\,\varepsilon_{a b c d}\, \delta_\sigma^d\,  \partial_{x^-}  \,\Bigl( \,\int_{-\infty}^{x^-}\,d \bar{x}^-\, \,E^c_-(x^+,\bar{x}^-, x_\bot)-\int^{\infty}_{x^-}\,d \bar{x}^-\, \,E^c_-(x^+,\bar{x}^-, x_\bot)\,\Bigr)^{ab}\, \delta^i_a \partial_{\bot,i} {\cal A}_{+}^{ab}(x^+,0,\vec{x}_\bot)\label{E22}
\end{eqnarray}
The fields coupled to the Wilson lines
$$\tilde{\cal A}^+(x^-,\vec{x}_\bot) = {\cal A}^+(0,x^-,\vec{x}_\bot),  \tilde{\cal A}^-(x^+,\vec{x}_\bot) = {\cal A}^-(x^+,0,\vec{x}_\bot)$$
satisfy the constraints
$$
\partial_\pm \tilde{\cal A}^\pm = \partial_\pm \tilde{\cal A}_\mp =0
$$
\revision{which guarantee, that $\tilde{\cal A}^\pm$ does not depend on $x^\pm$.
The same refers to
 $$\tilde{\cal E}^+(x^-,\vec{x}_\bot) = {\cal E}^+(0,x^-,\vec{x}_\bot),  \tilde{\cal E}^-(x^+,\vec{x}_\bot) = {\cal E}^-(x^+,0,\vec{x}_\bot)$$
which satisfy the constraints
$$
\partial_\pm \tilde{\cal E}^\pm = \partial_\pm \tilde{\cal E}_\mp =0
$$
 At the same time the kinetic term  contains $\cal A$ and $\cal E$ rather than $\tilde{\cal A}$ and $\tilde{\cal E}$.
We obtain that in the kinetic term  the field ${\cal A}^+(x^+,x^-,x_\bot)$ depends on $x^+$ while ${\cal A}^-(x^+,x^-,x_\bot)$ depends on $x^-$. It is worth mentioning, that the form of the action for QCD presented in \cite{LipatovEff,LipatovEff1,Our1} contains the kinetic term, in which the reggeon fields depend only on three coordinates: the "$-$" component depends on $x^+,x_\bot$ while the "$+$" component depends on $x^-,x_\bot$. } The experience of QCD \cite{Our3} shows that we may require that in kinetic term $\cal A$ and $\cal E$ obey the above constraints. In the following we will assume therefore ${\cal A} = \tilde{\cal A}$ and ${\cal E} = \tilde{\cal E}$.

\section{The time reversal invariant form of effective action and the classical equations of motion}

In the following calculations we use the simplified notations for the ordered exponentials. In these expressions indices are omitted typically, but they also may be written explicitly if necessary.
\begin{equation}\label{Ss7}
O(\omega_{\pm}) = {\Omega}_\pm(x^+,x^-,x_\bot)  = P\, {\rm exp}\,\Big(  \int_{-\infty}^{x^{\pm}} \omega_{\pm}(x)\, d x^{\pm} \Big)
\end{equation}
and
\begin{equation} \label{Ss8}
O^{T}(\omega_{\pm}) = \tilde{\Omega}_\pm(x^+,x^-,x_\bot)  = P\, {\rm exp}\,\Big(  \int^{\infty}_{x^{\pm}} \omega_{\pm}(x)\, d x^{\pm} \Big),
\end{equation}
We also denote
\begin{equation}
\Le \om_\mu\Ra^{a b }\,=\,\om_{\mu}^{\alpha \beta}\,\Le J_{\alpha \beta}\Ra^{a b}\,,
\end{equation}
by $J$ in the exponential we denote expression proportional to the generators of Lorentz group:
\beq\label{Addf4}
\Le J_{\alpha \beta}\Ra^{a b}\,=\,-\,\frac{1}{2}\,\Le \delta^a_\alpha\,\delta_{\beta}^{b}\,-\,
\delta^b_\alpha\,\delta_{\beta}^{a}\,\Ra\,.
\eeq

$S_{eff}=S^0[\om] + S_{ind}$ is an effective action for the interaction between the reggeons and virtual gravitons, and
\begin{eqnarray}\label{Ss501}
S_{ind}\,&=&\,-\,m_P^2\, \int d^4 x\,\varepsilon^{i - + \sigma}\,\varepsilon_{a b c d}\,\,\delta_\sigma^d\,\,
{\cal E}_{+}^{c}\,\,\partial_{i}\,{\cal A}_{-}^{a b }\,-
\,m_P^2\, \int d^4 x \,\varepsilon^{i + -  \sigma}\,\varepsilon_{a b c d}\,\,\delta_\sigma^d\,\,
\,{\cal E}_{-}^{c}\,\partial_{i}\, {\cal A}_{+}^{a b}\,-\,
\nonumber \\
&-&
\,m_P^2\, \int d^4 x\,\varepsilon^{i - + \sigma}\,\varepsilon_{a b c d}\,\delta_\sigma^d\,
\Le \partial_-\, O^{a b}(\om_{-}) \Ra\,\partial_{i}{\cal E}_{+}^{c}\,-\,
\,m_P^2\, \int d^4 x \,\varepsilon^{i + - \sigma}\,\varepsilon_{a b c d}\, \delta_\sigma^d\,\Le \partial_+\,
 O^{a b}(\om_{+}) \Ra\,\partial_{i}{\cal E}_{-}^{c}\,+\,
\nonumber \\
&+&
\,{m_P^2}\, \int d^4 x\,\varepsilon^{i - + \sigma}\,\varepsilon_{a b c d}\,\delta_\sigma^d\,\,
\Le\, e_{+}^{k}\,O_{k}\,^{c}(\om_{+})\,\Ra
\partial_{i}{\cal A}_{-}^{a b}\,+
\nonumber \\
&+&
\,{m_P^2}\, \int d^4 x \,\varepsilon^{i  + - \sigma}\,\varepsilon_{a b c d}\,\,\delta_\sigma^d\,\,
\Le\, e_{-}^{k}\,O_{k}\,^{c}(\om_{-})\,\Ra\,
\partial_{i}{\cal A}_{+}^{a b}\,.\label{E410}
\end{eqnarray}
We make the further  modification of \eq{E410}:
\begin{eqnarray}\label{Ss501}
S_{ind}\,&=&\,-\, m_P^2\, \int d^4 x\,\varepsilon^{i - + \sigma}\,\varepsilon_{a b c d}\,\,\delta_\sigma^d\,\,
{\cal E}_{+}^{c}\,\,\partial_{i}\,{\cal A}_{-}^{a b }\,-
\,m_P^2\, \int d^4 x \,\varepsilon^{i + -  \sigma}\,\varepsilon_{a b c d}\,\,\delta_\sigma^d\,\,
\,{\cal E}_{-}^{c}\,\partial_{i}\, {\cal A}_{+}^{a b}\,-\,
\nonumber \\
&-&
\,m_P^2\, \int d^4 x\,\varepsilon^{i - + \sigma}\,\varepsilon_{a b c d}\,\delta_\sigma^d\,
\Le \partial_-\, O^{a b}(\om_{-}) \Ra\,\partial_{i}{\cal E}_{+}^{c}\,-\,
\,m_P^2\, \int d^4 x \,\varepsilon^{i + - \sigma}\,\varepsilon_{a b c d}\, \delta_\sigma^d\,\Le \partial_+\,
 O^{a b}(\om_{+}) \Ra\,\partial_{i}{\cal E}_{-}^{c}\,+\,
\nonumber \\
&+&
\,\frac{m_P^2}{2}\, \int d^4 x\,\varepsilon^{i - + \sigma}\,\varepsilon_{a b c d}\,\delta_\sigma^d\,\,
\Le\, e_{+}^{k}\,O_{k}\,^{c}(\om_{+})\,+\,(O^{T})^c\,_k(\om_{+})\,\,e_{+}^{k}\,\Ra
\partial_{i}{\cal A}_{-}^{a b}\,+
\nonumber \\
&+&
\,\frac{m_P^2}{2}\, \int d^4 x \,\varepsilon^{i  + - \sigma}\,\varepsilon_{a b c d}\,\,\delta_\sigma^d\,\,
\Le\, e_{-}^{k}\,O_{k}\,^{c}(\om_{-})\,+\,(O^{T})^c\,_{k}(\om_{-})\,e_{-}^{k}\,\Ra\,
\partial_{i}{\cal A}_{+}^{a b}\,.\label{E41}
\end{eqnarray}
The first two rows of this expression repeat the first two rows of Eq. (\ref{E410}). The last two rows of Eq. (\ref{E410}) are modified. Namely, in the previous sections we considered basically the vierbein $e^i_b(x)$ transformed via parallel transporter that connects point $x$ with $x^{\pm} = -\infty$. In scattering theory there is no difference between $x^{\pm} = -\infty$ and $x^\pm = +\infty$, and we may repeat derivation of effective action taking as a reference $x^\pm = +\infty$ instead of $x^\pm = -\infty$, so that the field ${\cal E}^i_b$ will be transformed under the gauge transformations defined at $x^\pm = +\infty$. Both versions of effective theory are not invariant under time reversal. In the similar situation in QCD we proposed the form of Lipatov effective action that is explicitly invariant under the time reversal. Such an effective theory is constructed combining the two reference points $x^\pm = +\infty$ and $x^\pm = -\infty$. Similar combination has been applied above for the modification of the last two rows of Eq. (\ref{E410}), which results in the last two rows of Eq.  (\ref{E41}).

It is worth mentioning that looking at both Eq. (\ref{E41}) and Eq. (\ref{Scattering}) one may recognize the dominant contributions to the scattering amplitude as the ones coming from the propagators of $\cal E$ and $\cal A$. In fact, just as in the case of QCD we may replace in leading approximation the exchange by multiple quanta of $\omega$ and $e$ by the exchange by the single quanta of $\cal E$ and $\cal A$.


Now we can consider equations obtained via
variation of \eq{AddF1} and \eq{E41} with respect to $\om_{\pm}$. We have for the "bare" action
\begin{equation}
\delta_{\om_{\pm}}\,S\,=\,-m_{P}^{2}\,
\int\,d^4 x\,\varepsilon^{\mu \nu \rho \sigma}\,\varepsilon_{a b c d}\,\Le D_{\mu} e_{\rho}^{c}\Ra\,e_{\sigma}^{d}\,\Le \delta \,\om_{\nu}^{a b} \Ra \,\label{ClS1}
\end{equation}
In the absence of the induced part of the action extremum of the action is achieved at the value of spin connection that is given by its Riemannian expression through  $e_\rho^a$. Now we have the extra term in the action. Its variation is given by
\beqar\label{ClS82}
\delta_{\om_{\pm}}\,S_{ind}\,& = &\,-\,m_P^2\, \int d^4 x\,\varepsilon^{i \pm \mp d}\,\varepsilon_{a b c d}\,\Le  U^{a b}_{c_1 d_1}(\om_\pm) \,\delta\Le \om_{\pm}\Ra^{c_1 d_1}\,\Ra\,
\D_{i}{\cal E}^{c}_{\mp}\,+\,
\nonumber \\
&+&
\frac{m_P^2}{2} \int d^4 x^\prime \, \varepsilon^{i \mp \pm d}\,\varepsilon_{a b c d} \,
 \tilde{U}^{c}_{\alpha \beta r}(x^\pm,[x^\prime]^\pm )\,\delta \om_{\pm}^{\alpha \beta}(x^\prime) \,\partial_i {\cal A}_{\mp}^{a b} \int dx^\pm e_{\pm}^{k}\,O_{k}\,^{r}(\om_{\pm}) \nonumber \\
&+&
\frac{m_P^2}{2} \int d^4 x^\prime \, \varepsilon^{i \mp \pm d}\,\varepsilon_{a b c d} \,
 \tilde{U}^{Tc}_{\alpha \beta r}(x^\pm,[x^\prime]^\pm )\,\delta \om_{\pm}^{\alpha \beta}(x^\prime) \,\partial_i {\cal A}_{\mp}^{a b} \int dx^\pm \,O^{T r}\,_{k}(\om_{\pm})e_{\pm}^{k}
\,
\eeqar
In Appendix A it is shown that 
\begin{eqnarray}\label{Ss7_}
U^{b}_{\alpha \beta a}  &=&   \frac{1}{2}O^T(\omega_{\pm})_{a [\alpha}\,    O(\omega_{\pm})_{\beta]}\,^{b}  \nonumber\\&=&  \frac{1}{2} P\, {\rm exp}\,\Big(  \int^{+\infty}_{x^{\pm}} \omega_{\pm}(x)\, d x^{\pm} \Big) _{a[\alpha}  {\rm exp}\,\Big(  \int_{-\infty}^{x^{\pm}} \omega_{\pm}(x)\, d x^{\pm} \Big)_{\beta]}\,^{b}
\end{eqnarray}
Tensors $\tilde{U}$, $\tilde{U}^T$ may be calculated in the similar way. However, we actually do not need them because the corresponding terms may be neglected.
In our theory the finite cutoff is to be used for the integration over $x^\pm$. It appearance reflects the fact that the considered interactions occur during the short time distance of collision. We are left with the estimate
$$
 \int^{\infty}_{-\infty} {e}_{\pm}^{k}\, dx^{'\pm} \sim \frac{1}{\sqrt{s}}
$$
The term containing this factor gives negligible contribution to the scattering amplitudes in case of multi - Regge kinematics when $\sqrt{s} \gg \sqrt{|t|}$, and we omit them in classical equations of motion. This results in the following system of equations (for the details see Appendix A)
\begin{eqnarray}
&& \,
\Le D_{[j} e_{+]}^{[c} \Ra e_{-}^{d]} - \,\Le D_{[j} e_{-]}^{[c} \Ra e_{+}^{d]} +
\,\Le D_{[+} e_{-]}^{[c} \Ra e_{j}^{d]} \,=\,0\,.\label{KT0}\\&&
\,\Le D_{[i} e_{\mp]}^{[c}\Ra\,e^{d]}_{j}\,- \,\Le D_{[j} e_{\mp]}^{[c}\Ra\,e^{d]}_{i}- \,
\Le D_{[i} e_{j]}^{[d}\Ra\,e_{\mp}^{c]} \,= \frac{1}{2}\varepsilon^{abcd}\varepsilon_{a_1 b_1 c j}\,\Le  U^{a_1 b_1}_{a b}(\om_\pm) \,\Ra\,
\D_{i}{\cal E}^{c}_{\mp} -  \frac{1}{2} \varepsilon^{abcd}\varepsilon_{a_1 b_1 c i}\,\Le  U^{a_1 b_1}_{a b}(\om_\pm) \,\Ra\,
\D_{j}{\cal E}^{c}_{\mp}\,\nonumber
\end{eqnarray}
Here $i,j=1,2 $. 
This is the system of $24$ equations for $24$ components of spin connection. Notice also that the left hand sides of these equations are expressed through torsion tensor
 $T_{\nu \mu}^c =   D_{[\nu} e_{\mu ]}^{c}$ only.  
  Torsion has to be calculated via solution of Eqs. (\ref{KT0}). We do not give the precise solution of this equation, which is not important at this level of understanding of the theory. It is worth mentioning that counting the number of degrees of freedom allows to understand, that Eqs. (\ref{KT0}) fix all components of torsion unambiguously. In turn, the spin connection is given as a sum of Riemannian part and the contorsion. The latter is expressed through torsion.

Palatini action may be represented as follows
$$
S = - m_P^2 \int dx |e| (R -\frac{2}{3}T^2 + \frac{1}{24} S^2 + \frac{1}{2} q^2)
$$
Here $R$ is Riemannian curvature while $T,S$ and $q$ are the irreducible components of torsion (see Appendix A). Now variation of this action with respect to vierbein gives us the classical equations of motion for the latter. We should also take into account variation of the induced part of the action (the last two rows in Eq. (\ref{E41})). Altogether we obtain equations, in which the derivative of the field $\cal A$ plays the role of matter, i.e. the source of gravitational excitations.


At the next step we should develop perturbation theory. In order to avoid the overcounting of degrees of freedom we may develop it around the state with $\omega = 0$ and $e^a_i =\delta_i^a$. However, if we want to consider perturbations around classical vacuum with $\omega_{classical}[{\cal E},{\cal A}]$ and $e_{classical}[{\cal E},{\cal A}]$ given by the solutions of Eq. (\ref{KT0}) (and the corresponding classical equations for the vierbein), then we are to propose the scheme that allows to avoid over-counting of the degrees of freedom that lead to formation of the expected reggeons. Recall that the induced term of the action of Eq. (\ref{E41}) itself it obtained after the dimensional reduction of Palatini action. By construction it represents bare Palatini action taken for the particular configurations of gravitational fields corresponding to reggeons supplied by the interactions with gravitons. These configurations are assumed to be associated with the classical solutions for fixed $\cal A$ and $\cal E$.
In the similar situation in case of QCD we modified the kinetic term for the fields of reggeons. Now we propose the similar mechanism. More explicitly, we need to compose the effective action 
$$
S_{eff}[e,\omega,{\cal E},{\cal A}] = S_0[e,\omega] + S_{ind}[e,\omega,{\cal E},{\cal A}] + \Delta S_{ind}[{\cal E},{\cal A}]
$$
Here $S_0$ is Palatini action, $S_{ind}$ is given by Eq. (\ref{E41}) while $\Delta S_{ind}[{\cal E},{\cal A}]$ is chosen in such a way that
$$
S_{eff}\Big[e_{classical}[{\cal E},{\cal A}],\omega_{classical}[{\cal E},{\cal A}],{\cal E},{\cal A}\Big] = S_{0}\Big[e_{classical}[{\cal E},{\cal A}],\omega_{classical}[{\cal E},{\cal A}]\Big]
$$
For this purpose we chose
$$
\Delta S_{ind}[{\cal E},{\cal A}]=-S_{ind}\Big[e_{classical}[{\cal E},{\cal A}],\omega_{classical}[{\cal E},{\cal A}],{\cal E},{\cal A}\Big]
$$
This results in modification of kinetic term for the fields $\cal E$ and $\cal A$ in the effective action (the term that does not contain $e$ and $\omega$). 
If we would not make this modification of  the kinetic term for the reggeons, the perturbation theory developed around classical solutions for $e$ and $\omega$ depending on $\cal E$ and $\cal A$ would give the doubled $S^0$, which reflects that the degrees of freedom giving rise to (the expected) reggeons would be counted twice. In case of QCD \cite{Our3} this procedure leads to doubling of the kinetic term for the reggeons. In the present case the corresponding modification may be more valuable. 


\section{Conclusions and discussion}

The Regge Field Theory (RFT) calculus has a long history as a useful tool in the study of high-energy processes and calculation of corresponding amplitudes. It works reasonably well in an analysis of the high-energy amplitudes and the phenomenological description of experimental data (see \cite{0dim1,0dim2,0dim3,0dim4,Pom1,Pom2,Pom3}, for example). The idea of \cite{GravReg} to generalize Regge calculus to quantum gravity is natural. Modulo construction of the self consistent theory of quantum gravity it can allow us, in principle, to consider the scattering processes with gravitons exchange involved. The idea of L.N.Lipatov \cite{LipGrav}  was to write an effective action similarly to what is done in QCD in order to formulate the problem on the level of an effective Lagrangian with the reggeized gravitons included. It is worth mentioning that the effective action of \cite{LipGrav} is not completely similar to the one of
 \cite{LipatovEff}, where the same was done for the QCD RFT.

The main reason for that is that the ordinary Einstein gravity is not a gauge field theory with independent gauge field, and we cannot directly generalize the results of \cite{LipatovEff}
 to the case of Einstein gravity. In opposite, the Riemann-Cartan-Einstein formulation of gravity contains an independent gauge field of $SO(3,1)$ group.  This allows us to guess the $SO(3,1)$ gauge invariant effective action for the high - energy scattering. It includes interaction
of ordinary gravitational fields (vierbein and spin connection) with the fields that may be considered as candidates to the role of the reggeized counterparts of vierbein and spin connection.

First of all, based on heuristic arguments related to the dimensional reduction we propose bare form of effective action for the effective action of high energy scattering in Einstein gravity, see
\eq{Ch8} and \eq{Ss501}. The overall procedure is similar to the one proposed in \cite{Our3}. Namely, we consider the effective two - dimensional description of high energy scattering as a model operating with the Wilson lines along the light cone (corresponding to the $SO(3,1)$ connection) and translations along the light cone (corresponding to the integral of vielbein). The dimensional reasons and requirement of locality (in transverse coordinates) allow us to guess the particular form of $2D$ effective action (see also \cite{Balit}). Our candidates to the role of reggeon fields appear as a result of the Hubbard - Stratinovich transformation. The similar procedure performed in
\cite{Our3} (see also \cite{Verlinde}) has led to the Lipatov effective action for the high energy scattering in QCD.

In case of gravity the direct construction of this type fails to give the reasonable perturbation expansion in the resulting theory. It appears that the classical equations of motion in the presence of the reggeons contain unpleasant singularities. As a result the lowest order terms in perturbation theory do not work properly. The problem is solved if the cutoff is added to the integral along the light cone. Since we deal with the multi - regge kinematics the corresponding terms are proportional to   $\sim \frac{1}{\sqrt{s}}$. These contributions may be disregarded in the limit $s \gg |t|$.

It appears that due to
the interaction of ordinary fields with our new effective fields $\cal A$ and $\cal E$ (that are the reggeon candidates) torsion manifests itself in the considered theory already on the level of classical equations of motion. Let us recall that in the original theory without these effective fields torsion vanishes on the classical level.  The contortion tensor is determined by $\cal A$ and $\cal E$ (up to a certain ambiguity) if we solve equations of motion perturbatively. Still, we did not calculate directly the propagators of  $\cal A$ and $\cal E$, and did not clarify behavior of the simplest scattering amplitudes. Nevertheless,
we conclude that the effects of torsion in high energy scattering can be enhanced due the
exponential growth of the graviton propagator with energy. This is an additional important lesson of our calculations. In the effective theory of high energy scattering in Einstein - Cartan gravity we have the effects of  dynamical torsion due to the interaction of target and projectile.

Solution of the equations of motion, first for connection, and further for the vierbein, can be performed in the framework of perturbative scheme. One can solve equations of motion for both fields step by step providing necessary precision of the effective Lagrangian.
Concerning the intended graviton's reggeization, we notice that certain requirements for the reggeization are satisfied, see also discussion in \cite{Our1,Our2}. Nevertheless, the
reggeization in the effective action formalism can be directly proven only after the calculation of the corresponding propagators of  $\cal A$ and $\cal E$, which is not done yet in the present paper. In general,
this task can be performed similarly to what is done in \cite{Our1} for QCD. This task is postponed to the future work.

To conclude, we propose an effective action of  the high energy scattering in Einstein - Cartan gravity, which allows, in principle, to calculate various amplitudes with the exchange by multiple  gravitons. Construction of this effective theory has been performed on the basis of heuristic arguments related to the dimensional reduction, on the basis of natural requirements the needed theory is to satisfy, and on the basis of an analogy to the Lipatov's effective action in QCD.
Since we did not present the direct proof that the proposed effective action indeed describes high energy scattering in quantum gravity, it should be considered as a hypothesis to be checked by the further direct calculations. It is also worth mentioning that the quantum theory of gravity with Einstein (or Palatini) action does not exist. The simplest action linear in curvature has to be supplemented by various constraints and/or extra terms in the action in order to determine the well - defined quantum theory. Therefore, our construction is intended to be a part of a certain effective description of the true self - consistent gravity in its domain, where the term linear in curvature is relevant, while the extra terms in the action are not. Thus we are speaking about the energies of the processes much smaller than the Plank mass. This does not contradict, however, to the requirement that $s \gg |t|$ in these processes.

On of the authors (S.B.) kindly acknowledges useful discussions in the past with L.Lipatov. The authors are grateful to A.Onishchenko for careful reading of the manuscript and useful comments.

\appendix

\section{Classical equations of motion for torsion}

We consider equations obtained via
variation of \eq{AddF1} and \eq{E41} with respect to $\om_{\pm}$. We have for the "bare" action the variation given by \eq{ClS1}) while for the induced part of the action the variation is given in \eq{ClS8}).
Tensors entering these expressions are defined as follows
\beq\label{A5}
\delta \int dx^\pm \Le  \D_{\pm}  O(\om_{\pm})_{a}\,^{b}\,\Ra
\,=\, \int dx^\pm U^{b}_{\alpha \beta a}\,\delta \om_{\pm}^{\alpha \beta}
\eeq
while
\beq
\delta \int dx^\pm \Le\, e_{\pm}^{k}\,O_{k}\,^{c}(\om_{\pm})\,\Ra
\,=\, \int dx^\pm e_{\pm}^{k}\,O_{k}\,^{a}(\om_{\pm})  \int d[x^\prime]^\pm \tilde{U}^{c}_{\alpha \beta a}(x^\pm,[x^\prime]^\pm \,\delta \om_{\pm}^{\alpha \beta}(x^\prime)
\eeq
and
\beq
\delta \int dx^\pm \Le\, (O^{T})^c\,_k(\om_{\pm})\,\,e_{\pm}^{k}\,\Ra
\,=\, \int dx^\pm \Le\, (O^{T})^a\,_k(\om_{\pm})\,\,e_{\pm}^{k}\,\Ra  \int d[x^\prime]^\pm \tilde{U}^{T c}_{\alpha \beta a}(x^\pm,[x^\prime]^\pm) \,\delta \om_{\pm}^{\alpha \beta}(x^\prime)
\eeq
Tensor $U$ can be calculated as follows. We have
\begin{eqnarray}\label{Ss7__}
\delta O(\omega_{\pm})_{a}\,^{b}\Big|_{x^\pm = \infty} & = & \delta P\, {\rm exp}\,\Big(  \int^{+\infty}_{x^{\pm}} \omega_{\pm}(x)\, d x^{\pm} \Big)_{a}\,^{b} \nonumber\\ &=&  \delta \int dx^\pm   O^T(\omega_{\pm})_{a\alpha}\,  \delta \om_{\pm}^{\alpha \beta}  O(\omega_{\pm})_{\beta}\,^{b}  \nonumber\\&=& \int dx^\pm P\, {\rm exp}\,\Big(  \int^{+\infty}_{x^{\pm}} \omega_{\pm}(x)\, d x^{\pm} \Big) _{a\alpha}\delta \om_{\pm}^{\alpha \beta}   {\rm exp}\,\Big(  \int_{-\infty}^{x^{\pm}} \omega_{\pm}(x)\, d x^{\pm} \Big)_{\beta}\,^{b}
\end{eqnarray}
We obtain
\begin{eqnarray}\label{Ss7_}
U^{b}_{\alpha \beta a}  &=&   \frac{1}{2}O^T(\omega_{\pm})_{a [\alpha}\,    O(\omega_{\pm})_{\beta]}\,^{b}  \nonumber\\&=&  \frac{1}{2} P\, {\rm exp}\,\Big(  \int^{+\infty}_{x^{\pm}} \omega_{\pm}(x)\, d x^{\pm} \Big) _{a[\alpha}  {\rm exp}\,\Big(  \int_{-\infty}^{x^{\pm}} \omega_{\pm}(x)\, d x^{\pm} \Big)_{\beta]}\,^{b}
\end{eqnarray}
We rewrite the above mentioned expression for the variation of induced action as follows:
\beqar\label{ClS82A}
\delta_{\om_{\pm}}\,S_{ind}\,& = &\,-\,m_P^2\, \int d^4 x\,\varepsilon^{i \pm \mp d}\,\varepsilon_{a b c d}\,\Le  U^{a b}_{c_1 d_1}(\om_\pm) \,\delta\Le \om_{\pm}\Ra^{c_1 d_1}\,\Ra\,
\D_{i}{\cal E}^{c}_{\mp}\,+\,
\nonumber \\
&+&
\frac{m_P^2}{2} \int d^4 x^\prime \, \varepsilon^{i \mp \pm d}\,\varepsilon_{a b c d} \,
 \tilde{U}^{c}_{\alpha \beta a}(x^\pm,[x^\prime]^\pm )\,\delta \om_{\pm}^{\alpha \beta}(x^\prime) \,\partial_i {\cal A}_{\mp}^{a b} \int dx^\pm e_{\pm}^{k}\,O_{k}\,^{a}(\om_{\pm}) \nonumber \\
&+&
\frac{m_P^2}{2} \int d^4 x^\prime \, \varepsilon^{i \mp \pm d}\,\varepsilon_{a b c d} \,
 \tilde{U}^{Tc}_{\alpha \beta a}(x^\pm,[x^\prime]^\pm )\,\delta \om_{\pm}^{\alpha \beta}(x^\prime) \,\partial_i {\cal A}_{\mp}^{a b} \int dx^\pm \,O^{Ta}\,_{k}(\om_{\pm})e_{\pm}^{k}
\,
\eeqar
The second current in this expression contains contribution proportional to  $ \int^{\infty}_{-\infty} {e}_{\pm}^{k}\, dx^{\pm}$. Actually, in our theory the finite cutoff is to be used for the integration over $x^\pm$. It appearance reflects the fact that the considered interactions occur during the short time distance of collision. We are left with the estimate
$$
 \int^{\infty}_{-\infty} {e}_{\pm}^{k}\, dx^{\pm} \sim \frac{1}{\sqrt{s}}
$$
The term containing this factor gives negligible contribution to the scattering amplitudes in case of multi - Regge kinematics when $\sqrt{s} \gg \sqrt{|t|}$, therefore we do not take into account the
second current in \eq{ClS82} in our further calculations. This gives
\beqar\label{ClS82}
\delta_{\om_{\pm}}\,S_{ind}\,& = &-m_P^2\,\int\,d^4 x\,\Le\, \varepsilon^{i \pm \mp d}\,(j_{\,\om\,i\,\mp\,d})_{c_1\,d_1}\,\Ra\,\delta\Le \om_{\pm}\Ra^{c_1 d_1}\,.
\eeqar
where
$$
(j_{\,\om\,i\,\mp\,d})_{c_1\,d_1} = \varepsilon_{a b c d}\,\Le  U^{a b}_{c_1 d_1}(\om_\pm) \,\Ra\,
\D_{i}{\cal E}^{c}_{\mp}\,
$$

  In order to construct general solution for these equations we will consider the  complete set of equations, the first equation is obtained after the variation with respect to $\om_{\pm}$
\footnote{We denote here $[a\, b]\,=\,a\,b\,-\,b\,a\,$.}:
\beq\label{ClS1012}
\varepsilon^{i \pm \mp \sigma}\,\Le \varepsilon_{c_1 d_1 c d}\,\Le D_{[i} e_{\mp]}^{c}\Ra\,e_{\sigma}^{d}\,-\frac{1}{2}\,
\varepsilon_{c_1 d_1 c d}\,\Le D_{[i} e_{\sigma]}^{c}\Ra\,e_{\mp}^{d} \Ra\,=\,
\varepsilon^{i \pm \mp d}\,(j_{\,\om\,i\,\mp\,d})_{c_1\,d_1}\,.
\eeq

 The following system of equations can be obtained (here $i,j=1,2 $):
\beqar\label{ClS1013}
\varepsilon_{c_1 d_1 c d}\,\Le D_{[i} e_{\mp]}^{c}\Ra\,e^{d}_{j}\,- \varepsilon_{c_1 d_1 c d}\,\Le D_{[j} e_{\mp]}^{c}\Ra\,e^{d}_{i}- \varepsilon_{c_1 d_1 d c}\,
\Le D_{[i} e_{j]}^{d}\Ra\,e_{\mp}^{c} \,& =&
\,(j_{\,\om\,i\,\mp\,j})_{c_1\,d_1}\,- (j_{\,\om\,j\,\mp\,i})_{c_1\,d_1};
\eeqar
Another system of equations is obtained after variation with respect to $\om_{i}$ with $i=1,2$:
\beq\label{ClS12}
\varepsilon^{\mu i \rho  \sigma}\,\varepsilon_{a b c d}\,\Le\,D_{\mu} e^{c}_{\rho}\,\Ra\,e^{d}_{\sigma}\,\delta \Le \om_{i} \Ra^{a b} \,=\,0\,
\eeq
which can be also written as
\beq\label{ClS14}
\varepsilon^{i j + -} \varepsilon_{a b c d }\,\Le
\Le D_{[j} e_{+]}^{c} \Ra e_{-}^{d} - \Le D_{[j} e_{-]}^{c} \Ra e_{+}^{d} +
\Le D_{[+} e_{-]}^{c} \Ra e_{j}^{d} \Ra\,=\,0\,.
\eeq
In order solve \eq{ClS1013} and \eq{ClS14} simultaneously we rewrite \eq{ClS14} in a different form and come to the system of equations:
\begin{eqnarray}\label{ClS15}
&& \varepsilon_{a b c d }\,
\Le D_{[j} e_{+]}^{c} \Ra e_{-}^{d} - \varepsilon_{a b c d }\,\Le D_{[j} e_{-]}^{c} \Ra e_{+}^{d} +
\varepsilon_{a b c d }\,\Le D_{[+} e_{-]}^{c} \Ra e_{j}^{d} \,=\,0\,.\nonumber\\&&
\varepsilon_{a b c d}\,\Le D_{[i} e_{\mp]}^{c}\Ra\,e^{d}_{j}\,- \varepsilon_{a b c d}\,\Le D_{[j} e_{\mp]}^{c}\Ra\,e^{d}_{i}- \varepsilon_{a b d c}\,
\Le D_{[i} e_{j]}^{d}\Ra\,e_{\mp}^{c} \,= \varepsilon_{a_1 b_1 c j}\,\Le  U^{a_1 b_1}_{a b}(\om_\pm) \,\Ra\,
\D_{i}{\cal E}^{c}_{\mp} -  \varepsilon_{a_1 b_1 c i}\,\Le  U^{a_1 b_1}_{a b}(\om_\pm) \,\Ra\,
\D_{j}{\cal E}^{c}_{\mp}\,\nonumber
\end{eqnarray}
This expression may be simplified further:
\begin{eqnarray}
&& \,
\Le D_{[j} e_{+]}^{[c} \Ra e_{-}^{d]} - \,\Le D_{[j} e_{-]}^{[c} \Ra e_{+}^{d]} +
\,\Le D_{[+} e_{-]}^{[c} \Ra e_{j}^{d]} \,=\,0\,.\label{KT}\\&&
\,\Le D_{[i} e_{\mp]}^{[c}\Ra\,e^{d]}_{j}\,- \,\Le D_{[j} e_{\mp]}^{[c}\Ra\,e^{d]}_{i}- \,
\Le D_{[i} e_{j]}^{[d}\Ra\,e_{\mp}^{c]} \,=  \frac{1}{2}\varepsilon^{abcd}\varepsilon_{a_1 b_1 c j}\,\Le  U^{a_1 b_1}_{a b}(\om_\pm) \,\Ra\,
\D_{i}{\cal E}^{c}_{\mp} -  \frac{1}{2} \varepsilon^{abcd}\varepsilon_{a_1 b_1 c i}\,\Le  U^{a_1 b_1}_{a b}(\om_\pm) \,\Ra\,
\D_{j}{\cal E}^{c}_{\mp}\,\nonumber
\end{eqnarray}
The left hand sides of these equations are expressed through torsion tensor
 $T_{\nu \mu}^c =   D_{[\nu} e_{\mu ]}^{c}$ only.  We also denote
$$
T^{\rho}_{\, . \,\nu \mu} =g^{\rho \sigma} e^d_\sigma \eta_{dc} T^{c}_{\nu \mu}
$$
It is instructive to introduce the irreducible components of torsion:
$$
T_\beta = T^{\alpha}_{\, . \, \beta\alpha}, \quad S^\nu = \epsilon^{\alpha \beta \mu \nu} T_{\alpha \beta \mu}
$$
and
$$
q_{\alpha \beta \mu} = T_{\alpha\beta \mu} + \frac{1}{6} \epsilon_{\alpha \beta \mu \nu} S^\nu - \frac{1}{3} (T_\beta g_{\alpha \mu} - T_\mu g_{\alpha\beta})
$$
One can check, that the reggeon field $\cal E$ appears to be the source of the linear combination of all three irreducible components of torsion.

Introducing the inverse vierbein, $E_{a}^{\mu}\,e^{b}_{\mu}\,=\,\delta^{b}_{a}\,$, we obtain the general form for the solution of  equations (\ref{KT}):
\beq\label{ps6}
\om_{\mu}^{a b}\,=\,\gamma_{\mu}^{a b}\,+\,K_{\mu}^{a b }\,.
\eeq
Here $\gamma_\mu^{ab} $ is the torsionless spin connection of general relativity while $K$ is the so - called contorsion tensor expressed through the above introduced torsion:
\beqar\label{ps7}
\gamma_{\mu}^{a b}\,& = &\,e_{\mu}^{c}\,\Le\,C_{c}\,^{a b} \,-\, C^{a b}\,_{c}\, + \,C^{b}\,_{c}\,^{a}\,\Ra\,/\,2, \\
K_{\mu}^{a b }\,& = & \,-\,e_{\mu}^{c}\,\Le\,T_{c}\,^{a b}\, -\, T^{a b}\,_{c}\, + \,T^{b}\,_{c}\,^{a}\,\Ra\,/\,2\,, \\
C_{a b}\,^{c}\,&=&\,E_{a}^{\mu}\,E_{b}^{\nu}\, \D_{[\mu} e_{\nu]}^{c}\,,
T_{a b}\,^{c}\,=\,E_{a}^{\mu}\,E_{b}^{\nu}\,T_{\mu \nu}\,^{c}\,\,,\,\,\,
\eeqar
Torsion has to be calculated via solution of Eqs. (\ref{KT}). We do not give the precise solution of this equation, which is not important at this level of understanding of the theory.

\end{document}